\documentclass[aps,prl,twocolumn,superscriptaddress,showpacs]{revtex4-1}

\usepackage{epsf}
\usepackage{graphicx}
\usepackage{amssymb}
\usepackage{color}

\newcommand{\SrIrO}{Sr$_2$IrO$_4$}

\newcommand{\LCO}{La$_2$CuO$_4$}

\newcommand{\Tn}{$T_{\rm N}$}

\begin{document}

\title{Low energy magnetic excitations in the spin-orbital Mott insulator \SrIrO}

\author{S.\ Bahr}
\affiliation{Leibniz Institute for Solid State and Materials Research IFW Dresden,
D-01171 Dresden, Germany}

\author{A.\ Alfonsov}
\affiliation{Leibniz Institute for Solid State and Materials Research IFW Dresden,
D-01171 Dresden, Germany}

\author{G.\ Jackeli}
\affiliation{Max Planck Institute for Solid State Research, Heisenbergstrasse 1, D-70569
Stuttgart, Germany}

\author{G.\ Khaliullin}
\affiliation{Max Planck Institute for Solid State Research, Heisenbergstrasse 1, D-70569
Stuttgart, Germany}

\author{A. Matsumoto}
\affiliation{Department of Physics and Department of Advanced Materials, University of Tokyo, Hongo 113-0033, Japan}

\author{T. Takayama}
\affiliation{Max Planck Institute for Solid State Research, Heisenbergstrasse 1, D-70569 Stuttgart, Germany}

\author{H.\ Takagi}
\affiliation{Department of Physics and Department of Advanced Materials, University of Tokyo, Hongo 113-0033, Japan}\affiliation{Max Planck Institute
for Solid State Research, Heisenbergstrasse 1, D-70569 Stuttgart, Germany}

\author{B.\ B\"{u}chner}
\affiliation{Leibniz Institute for Solid State and Materials Research IFW Dresden,
D-01171 Dresden, Germany} \affiliation{ Institut f\"{u}r Festk\"{o}rperphysik, Technische
Universit\"{a}t Dresden, D-01062 Dresden, Germany}

\author{V.\ Kataev}
\affiliation{Leibniz Institute for Solid State and Materials Research IFW Dresden,
D-01171 Dresden, Germany}

\date{\today}

\begin{abstract}

We report a high-field electron spin resonance study in the sub-THz
frequency domain of a single crystal of \SrIrO\ that has been recently
proposed as a prototypical spin-orbital Mott insulator. In the
antiferromagnetically (AFM) ordered state with noncollinear spin structure
that occurs in this material at \Tn\,$\approx$\,240\,K we observe both the
"low" frequency mode due to the precession of weak ferromagnetic moments
arising from a spin canting, and the "high" frequency modes due to the
precession of the AFM sublattices. Surprisingly, the energy gap for the
AFM excitations appears to be very small, amounting to 0.83\,meV only.
This suggests a rather isotropic Heisenberg dynamics of interacting
Ir$^{4+}$ effective spins despite the spin-orbital entanglement in the
ground state.

\pacs{75.30.-m, 75.50.Ee, 76.30.-v}

% 75.30.-m    Intrinsic properties of magnetically ordered materials

% 75.50.Ee    Antiferromagnetics

% 76.30.-v    Electron paramagnetic resonance and relaxation

\end{abstract}

\maketitle

\begin{figure}
\includegraphics[width=0.8\columnwidth]{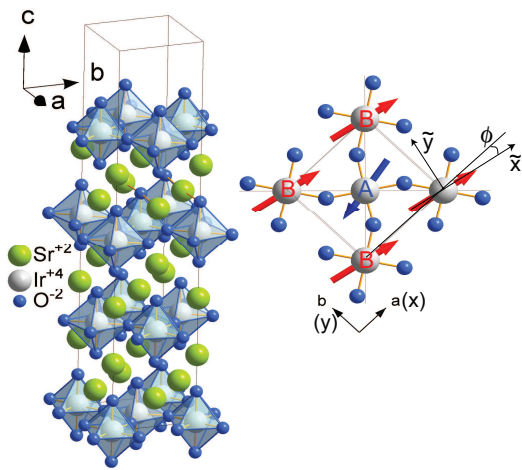}
\caption{(Color online) Left: crystallographic unit cell of \SrIrO. Right: the ordering pattern of effective Ir spins $J_{\rm eff}=1/2$ in the $ab$
crystallographic plane \cite{Kim09}. The spin in the center and its four neighbors belong to the two AFM sublattices, $A$ and $B$ respectively,
%, which are
canted by an angle $\phi$.} \label{structure}
\end{figure}

Many 3$d$ transition metal oxides (TMOs) such as parent compounds of the cuprate high-temperature superconductors are insulators despite a partial
filling of the $d$ shell. The Mott (or charge transfer) gap of the order of several eV opens in the 3$d$ band  due to a strong on-site Coulomb
repulsion of electrons $U$, rendering these materials magnetic insulators (for a review see, e.g., Ref.~\cite{Imada98}). In 5$d$ TMOs the $d$ shell
is more extended in space yielding a stronger overlap of the orbitals which favors electron delocalization and reduces $U$. Nevertheless, an
insulating magnetic ground state still often occurs, e.g., in complex Ir oxides. One striking example is \SrIrO\  which is analogous (see
Fig.~\ref{structure}) from the viewpoint of crystallographic \cite{Crawford94} and magnetic structure \cite{Kim09} to the celebrated two-dimensional
(2D) spin-1/2 Heisenberg quantum antiferromagnet \LCO. A recent study of the electronic structure of \SrIrO\  \cite{Kim08} has shown that the
insulating state arises due to a strong spin-orbit coupling (SOC) of the order of 0.5\,eV that gives rise to a narrow half-filled band characterized
by the effective total angular momentum $J_{\rm eff} = 1/2$. Even a small $U$ opens the Mott gap of $\sim 0.5$\,eV in this band \cite{Kim08,Moon09}.
In contrast to the cuprates, the driving force for the insulating state in \SrIrO\ is the SOC which suggests this material as a novel kind of
spin-orbital Mott insulator \cite{Kim08,Kim09}. On the atomic level, a coupling of the spin and orbital momentum in Ir$^{4+}$ ion yields a Kramers
doublet with the effective spin $J_{\rm eff} = 1/2$ which however has very different properties as compared to the real spin-1/2 state in \LCO.
Recent theories suggest that the entanglement of the spin and orbital momentum in Ir oxides can lead to a rich diversity of magnetic ground states,
ranging from a "conventional" N\'eel antiferromagentic (AFM) state, to stripe AFM phases and to a spin-liquid regime
\cite{Jackeli09,Chaloupka10,Balents14}. In \SrIrO\ the $J_{\rm eff}$ spins AFM order in the $ab$ plane below \Tn\,$\approx 240$\,K. Owing to the
admixture of the orbital moment the effective spins follow the rotation of the IrO octahedra which yields noncollinearity of the spin sublattices
\cite{Kim09} (Fig.~\ref{structure}). The value of the in-plane ferromagnetic (FM) moment $M_{\rm FM}\sim 0.06-0.1\mu_{\rm B}$/Ir
\cite{Crawford94,Cao98,Kim09,Fujiyama12}, which arises due to the Dzyaloshinskii-Moriya (DM) interaction, is much larger than in \LCO, as expected
because of the strong SOC. Surprisingly, dynamic correlations of $J_{\rm eff}$ spins in the paramagnetic state of \SrIrO\ reveal isotropic 2D
Heisenberg behavior despite the strong entanglement of spins and orbitals \cite{Fujiyama12}.

The low energy dynamics of effective spins in \SrIrO\ below \Tn\ has not been yet sufficiently addressed. Inelastic neutron scattering experiments
are complicated by a strong absorption of neutrons by Ir and, to our knowledge, have not been reported so far. The magnon dispersion has been
recently measured by the resonant inelastic x-ray scattering (RIXS) \cite{Kim12}. However a limited energy resolution did not enable to resolve a
possible gap in the excitation spectrum which could arise due to the SOC. In this situation electron spin resonance (ESR) spectroscopy, with its very
high energy resolution down to $\sim \mu$eV and extreme sensitivity to magnetic anisotropies, could provide fundamental insights into properties of
the magnetic excitations in spin-orbital Mott insulators. In the present work we study the low-energy spin dynamics in single crystals of \SrIrO\ in
the AFM ordered state with the ESR technique in a broad frequency range up to 500\,GHz. We have identified both the ferromagnetic-like resonance
mode, associated with the precession of the DM moments, and the AFM resonance (AFMR) modes due to the resonant excitation of the AFM sublattices. The
latter modes reveal a very small spin wave gap $\Delta = 0.83$\,meV which is just a fraction of that observed in \LCO. Indeed, it would not be
possible to resolve such a small gap in the RIXS experiment \cite{Kim12}. Our observation of practically gapless excitations of effective $J_{\rm
eff}$ spins in the spin-orbital Mott insulator is remarkable. It appears that, in the magnetic sector, despite the spin-orbital entanglement \SrIrO\
shows the behavior of a quasi-2D Heisenberg antiferromagnet.

The growth and characterization of single crystals of \SrIrO\ have been described in Ref.~\cite{Kim09}. ESR measurements have been performed with a
commercial Bruker EMX spectrometer  at a fixed frequency $\nu\sim 10$\,GHz, at fields up to 1\,T and with a home-made multi-frequency high-field ESR
(HF-ESR) spectrometer  at fields up to 14\,T and $\nu$ up to 500\,GHz \cite{Golze06}. In the latter case a highly sensitive quasi-optical detection
scheme in the so-called induction mode without resonance cavities has been employed \cite{Schaufuss09,Fuchs99}.

In the AFM ordered IrO$_2$ plane of \SrIrO\  a net ferromagnetic moment arises due to the DM interaction. These FM (or DM) moments are AFM coupled
along the $c$~axis in the up-down-down-up ($\uparrow\downarrow\downarrow\uparrow$) arrangement \cite{Kim09}. Application of the in-plane magnetic
field above the critical value $H_{\rm c}$ yields a metamagnetic transition from the AFM to a weakly FM state with $\mu_0H_{\rm c}$ increasing up to
$\sim 0.2$\,T at low temperatures \cite{Cao98,Kim09}. Therefore, as is generally expected for a canted two-sublattice antiferromagnet with "hidden"
ferromagnetism, there should be a low-frequency ferromagnetic resonance (FMR) mode that represents the oscillation of the net FM moment around its
equilibrium position and a high-frequency mode that arises due to the precession of the AFM sublattices \cite{Pincus60,Fink63,Williamson64,Turov65}.
Though qualitatively similar to the case of collinear AFM lattices, the latter resonance mode is renormalized by the DM interaction (see Refs.
\cite{Pincus60,Fink63,Williamson64,Turov65} and the discussion below).

\begin{figure}
\includegraphics[width=0.9\columnwidth]{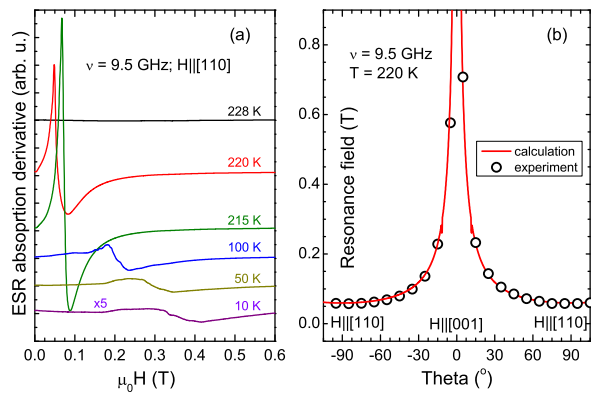}
\caption{(Color online) (a)  FMR signal at $\nu = 9.5$\,GHz at selected temperatures below \Tn\ for the in-plane orientation of the magnetic field;
(b)  orientational dependence of the FMR resonance field at $\nu = 9.5$\,GHz and $T = 220$\,K (circles). Solid line shows $1/\sin\theta$ fit  (see text).} \label{FMR}
\end{figure}

Experimentally, we observe the FMR mode in \SrIrO\ at fields above $H_{\rm
c}$ by measuring the ESR with a 10\,GHz setup. As illustrated by
Fig.~\ref{FMR}(a) for the magnetic field geometry ${\bf H}||[110]$, the
signal appears right below the AFM phase transition. (Note that in this
setup the detected signal is the field derivative of the microwave
absorption). The resonance field $H_{\rm res}$ shifts to higher fields
with lowering the temperature, the resonance line first grows in
amplitude, however, below $\sim 100$\,K the signal substantially broadens
and acquires an additional structure. In classical FMR theories the
increase of $H_{\rm res}$ should be related with a decrease of the
effective DM field which enters the resonance conditions and is
responsible for the downshift of the FMR mode from the paramagnetic
resonance value \cite{Turov65}. Interestingly,  after an initial rise
below \Tn\  the in-plane magnetization due to DM moments also decreases
upon lowering the $T$ at fields even exceeding $H_{\rm c}$, as observed in
our magnetization measurements (not shown) and also previously reported,
e.g., in Ref.~{\cite{Chicara09}}. Furthermore, a distribution of internal
fields in \SrIrO\ has been observed below $\sim 100$\,K in $\mu$SR
experiments in Ref.~\cite{Franke11}. It could be presumably assigned to
the pinning of the DM moments by defects and be the reason for the
broadening and structuring of the FMR mode at low $T$ [Fig.~\ref{FMR}(a)].

The rotation of the magnetic field from the in-plane $ab$ orientation towards the $c$ axis yields a strong shift of the FMR line to higher fields
[Fig.~\ref{FMR}(b)]. The angular dependence of resonance field at constant frequency follows  perfectly a well known $1/\sin\theta$ law [solid line
in Fig.~\ref{FMR}(b)] characteristic for the low-frequency FMR mode \cite{Pincus60}.

\begin{figure}
\includegraphics[width=0.8\columnwidth]{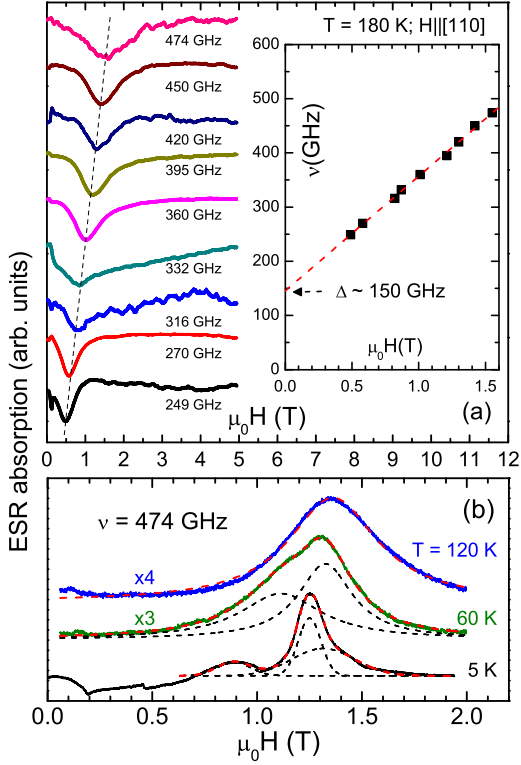}
\caption{(Color online) High-frequency AFMR mode for the in-plane orientation of the magnetic field. Upper panel: Characteristic frequency dependence
at $T = 180$\,K. The $\nu(H)$ AFMR resonance branch shown in the inset reveals a gap $\Delta\approx 150$\,GHz. Bottom panel: Selected spectra at
several temperatures for a fixed frequency $\nu = 474$\,GHz. Dashed lines are the Lorentzian fits.} \label{AFMR}
\end{figure}

\begin{figure}
\includegraphics[width=0.9\columnwidth]{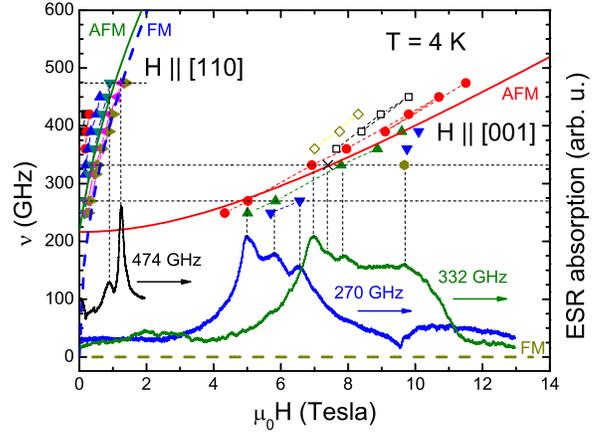}
\caption{(Color online) $\nu(H)$ diagram of the AFMR excitations in
\SrIrO\ at $T = 4 $\,K together with characteristic spectra for two
directions of the magnetic field [symbols - experimental data points,
solid lines - result of the modeling according to Eqs.~(\ref{eq2}) and
(\ref{eq3}). Thick dashed lines show the corresponding FMR modes. (see the
text)]} \label{diagram}
\end{figure}

The central result of our work is the observation of high-frequency AFMR modes in the HF-ESR measurements below \Tn. Compared to the FMR signal, the
AFM resonances are shifted up into the sub-THz frequency domain. As an example the AFMR signals at $T = 180$\,K and ${\bf H}||[110]$ are shown in
Fig.~\ref{AFMR}(a) for several selected frequencies. This mode reveals a very steep $\nu(H)$ dependence with a frequency cutoff $\nu (H=0) =
150$\,GHz [Fig.~\ref{AFMR}(a), inset]. Similar to the behavior of the FMR mode, lowering the temperature below $\sim 100$\,K yields the structuring
of a single Lorentzian line into a group of overlapping resonances [Fig.~\ref{AFMR}(b)], again suggesting some distribution of internal fields in the
sample. Nevertheless, all these closely lying resonances disperse in magnetic field very similar way and form a common $\nu(H)$ branch of excitations
with an energy gap $\Delta \approx 200$\, GHz (=\,0.83\,meV) at $T = 4$\,K (Fig.~\ref{diagram}). The AFM mode for ${\bf H}||{\bf c}$ also acquires
structure at low $T$ but reveals a much shallower $\nu(H)$ dependence (Fig.~\ref{diagram}). However, this AFM $c$-axis branch can be extrapolated to
the same frequency cutoff $\Delta$ at zero field as the $ab$-plane branch.

With ESR one probes the AFM excitations in the center of the magnetic Brillouin zone, and the excitation gap $\Delta$ corresponds to the spin wave
(SW) gap at the $q = 0$ wave vector \cite{Turov65}. We discuss the origin of
this gap and its field dependencies based on
an effective spin $S=1/2$ model suggested for \SrIrO\ in
Ref.~\cite{Jackeli09}. The model describes exchange interactions on
intra-layer bonds of neighboring iridium ions and has the form:
\begin{eqnarray}
{\cal H}_{ij}=J\vec S_i\cdot \vec S_{j}+\Gamma S_{i}^{z}S_{j}^{z} +D{\big (}S_i^xS_j^y-S_i^yS_j^x{\big )}~. \label{eq1}
\end{eqnarray}
Here the first term  stands for  isotropic AFM exchange ($J$), second and
third terms describe symmetric ($\Gamma$) and antisymmetric ($D$) exchange
anisotropies, respectively.  We derive resonance frequencies within the
linear SW theory. For the applied field $H$ along the $c$-axis, the
low-frequency FMR mode is gapless and field independent, and the
high-frequency AFMR one is at energy
\begin{equation}
h\nu_{c}=\sqrt{32{\tilde J}{\tilde \Gamma}S^2+(g_c\mu_BH)^2/(1-{\tilde \Gamma}/2{\tilde
  J})} ~.
\label{eq2}
\end{equation}
Here, ${\tilde J}=\sqrt{J^2+D^2}$ and ${\tilde \Gamma}={\tilde J}-J-\Gamma>0$ describe effective isotropic exchange
and easy-plane anisotropy, respectively.

When the applied field is in $ab$-plane, both modes are gapped. Retaining the leading order contribution in $H$,  we find the energy
of AFMR mode
\begin{equation}
h\nu_{ab}\approx\sqrt{32{\tilde J}{\tilde \Gamma}S^2+4(2{\tilde J}+{\tilde \Gamma})M_{\rm FM}H} \label{eq3}~,
\end{equation}
where  $M_{\rm FM}=Sg_{ab}\mu_B\sin\phi $ is in-plane FM moment due to
spin canting induced by DM interaction,   and $\tan2\phi=D/J$. The
corresponding FMR mode $[(g_{ab}\mu_BH)^2+4(2{\tilde J}+{\tilde
\Gamma})M_{\rm FM}H]^{1/2}$ lies, in the high frequency domain, close to
the AFMR one (Fig.~\ref{diagram}). Considering the broadness of the high
frequency ESR spectrum which at low $T$ also features  several overlapping
resonances, it is difficult to distinguish there an FMR signal.

With ${\tilde J}\approx 100$~meV, as estimated in Ref.~\cite{Fujiyama12}, and with the SW gap value $\Delta(H=0) \approx$ \,0.83\,meV (Fig.~\ref{diagram}) one
obtains from Eqs.~(\ref{eq2}) and (\ref{eq3}) a surprisingly small value of the anisotropy exchange parameter ${\tilde \Gamma} = \Delta^2/8J \approx
1$\,$\mu$eV. Since ${\tilde J}\gg {\tilde \Gamma}$, the slope of the $c$ axis AFMR $\nu(H)$ branch [Eq.~\ref{eq2})] in fields $H > \Delta/g_c\mu_{\rm B}$  is given
mainly by the $g$ factor. Therefore one can obtain the $g_c$ value directly
from the experimental data in Fig.~\ref{diagram} which yields $g_c\simeq 2.4$. A plot of Eq.~(\ref{eq2}) with these parameters is in a fairly good agreement with the experiment (Fig.~\ref{diagram}).
For the $ab$ plane such direct determination of the $g$ factor is not possible since the slope of the corresponding branch is strongly renormalized
by the DM interaction induced in-plane FM moment $M_{\rm FM}$, giving rise to a very steep $\nu(H)$ dependence [see Eq.~(\ref{eq3}) and Fig.~\ref{diagram}].
With  $M_{\rm FM}\approx 0.06\mu_{\rm B}$ \cite{Fujiyama12} and the parameters estimated above, Eq.~(\ref{eq3}) gives satisfactory agreement with experimental data [Fig.~\ref{diagram}].

The  effective easy-plane anisotropy ${\tilde\Gamma}$, extracted above, appears to be surprisingly small, suggesting  that the effects of  the DM
interaction $D$ [see Eq.~(\ref{eq1})], favoring in-plane  canted spin order, and easy-axis anisotropy $\Gamma$, supporting collinear out-of-plane
order, nearly compensate each other.  This is a plausible scenario,  requiring, however, ``fine-tuning'' of model parameters  \cite{Jackeli09}.
Alternatively, one may  attribute the emergent tiny energy scale ${\tilde \Gamma}$ to  weak inter-layer interactions and the observed modes to the
antibonding states derived from single-layer in-plane FMR modes. However, within this second scenario we were not able to explain the field
dependence of resonance frequency for  ${\bf H}||{\bf c}$, see Fig.~\ref{diagram}, as for this geometry single-layer in-plane FMR mode is gapless and
field independent. Since the DM moments are confined to the basal plane the magnetic field dependence of the FMR mode is governed by the transverse
component of the field $H_\perp = H\sin{\theta}$ only  and vanishes in the parallel geometry \cite{Pincus60,Fink63,Turov65}. For ${\bf H}||{\bf c}$,
a weak inter-layer coupling $J^\prime \ll J$ may only bring a weak field dependence of the modes derived from the in-plane FMR mode $h\nu_{c}^{\rm
FM} \sim (J^\prime/J)g_c\mu_{\rm B}H$, contrary to the observed  nearly paramagnetic  $h\nu_{c}^{\rm AFM}\sim g_c\mu_{\rm B}H$ behavior, as indeed
expected from Eq.~(\ref{eq2}).

Our observation of the AFM excitations in \SrIrO\ down to quite small frequencies despite the  spin-orbital entangled nature of ground state doublet
of iridium ion is striking. One would expect that an orbital contribution should, instead, give rise to a substantial magnetic anisotropy gap for the
spin excitations. For comparison, the SW gap in the AFM state of \LCO\ with a very similar magnetic structure and a similar exchange $J$ amounts to
$\Delta_{\rm LCO}\sim 5$\,meV (see., e.g., Ref.~\cite{Keimer93}). This copper oxide is considered to be nearly ideal realization of the Heisenberg S
= 1/2 2D quantum antiferromagnet due to an almost complete quenching of the orbital momentum of the Cu, and yet $\Delta_{\rm LCO}$ appears to be
several times larger than the gap $\Delta(H = 0) = 0.83$\,meV in \SrIrO. A similarly small gap has been observed by AFMR in another Cu-based canted
quasi-2D Heisenberg antiferromagnet Cu(HCOO)$_2\cdot$4H$_2$O \cite{Seehra70} where, however, the isotropic $J\approx 72$\,K is by a factor $\sim 15 -
20$ smaller than in \LCO\ and \SrIrO. Given that the ratio $\Delta/J$ is usually thought to increase with SOC, it is really surprising that $\Delta/J
\sim 0.008$ in strong spin-orbit coupled \SrIrO\ is much smaller than in weak SOC cuprates \LCO ($\Delta/J \sim 0.04$) and Cu(HCOO)$_2\cdot$4H$_2$O
($\Delta/J \sim 0.07$). Remarkably, a recent resonant magnetic x-ray scattering study of \SrIrO\ has revealed essentially isotropic Heisenberg-like
behavior of dynamic magnetic correlations above \Tn\ \cite{Fujiyama12} quite similar to the behavior of \LCO\ \cite{Keimer92}. It is therefore
plausible that below \Tn\ such isotropic magnetic correlations result in a rather isotropic magnetic dynamics in \SrIrO\ despite a strong
entanglement of spin and orbital degrees of freedom.  One should note a sizable  antisymmetric DM anisotropy in \SrIrO\ that leads to a "weak" FM
moment of two orders of magnitude larger than that in \LCO. However, the competition between antisymmetric and symmetric anisotropies (both being
induced by the strong spin-orbit coupling and being sizable) may lead to  a quasi-degeneracy of in-plane canted and out-of-plane collinear orders,
thereby explaining an extremely small SW gap. The collinear $c$~axis magnetic structure has indeed been observed in bilayer iridate Sr$_3$Ir$_2$O$_7$
\cite{Kim12a} and in Sr$_2$Ir$_{0.9}$Mn$_{0.1}$O$_4$ \cite{Calder12}.

In summary, we have studied by means of ESR the low energy dynamics of
effective $J_{\rm eff}$ spins in the magnetically ordered state of the
prototypical spin-orbital Mott insulator \SrIrO. The observed resonance of
the DM moments  with its angular dependence can be well described by a
standard phenomenological treatment of ferromagnetic resonance.  In the
sub-THz frequency domain we identify the resonance modes due to the
oscillation of the AFM sublattices. Surprisingly, the AFMR branches reveal
a very small spin wave gap at $H = 0$ amounting to $\sim 0.83$\,meV only,
far beyond the resolution limit of RIXS \cite{Kim12}. The smallness of the
gap enables to classify \SrIrO\ as an isotropic quasi-2D Heisenberg
antiferromagnet, despite the spin-orbital entanglement in the ground
state. Considering experimental constraints of the inelastic neutron
scattering (strong neutron absorption) and RIXS (limited energy
resolution), high-field high-frequency tunable ESR technique with its
excellent energy resolution and very high sensitivity to magnetic
anisotropies emerges as an instructive tool to study novel magnetic phases
and low energy excitations in complex iridium based oxides.

This work has been supported in part by the Deutsche Forschungsgemeinschaft (DFG) through project FOR 912 "Coherence and relaxation of electron
spins".

\bibliography{literature_ESR_Sr2IrO4}

\end{document}